\documentclass[]{article}
\input epsf

\voffset= - 1.0in
\hoffset= - 1.0in         

\catcode`\@=11
\def\marginnote#1{}

\newcount\hour
\newcount\minute
\newtoks\amorpm
\hour=\time\divide\hour by60
\minute=\time{\multiply\hour by60 \global\advance\minute by-\hour}
\edef\standardtime{{\ifnum\hour<12 \global\amorpm={am}%
        \else\global\amorpm={pm}\advance\hour by-12 \fi
        \ifnum\hour=0 \hour=12 \fi
        \number\hour:\ifnum\minute<10 0\fi\number\minute\the\amorpm}}
\edef\militarytime{\number\hour:\ifnum\minute<10 0\fi\number\minute}

\def\draftlabel#1{{\@bsphack\if@filesw {\let\thepage\relax
   \xdef\@gtempa{\write\@auxout{\string
      \newlabel{#1}{{\@currentlabel}{\thepage}}}}}\@gtempa
   \if@nobreak \ifvmode\nobreak\fi\fi\fi\@esphack}
        \gdef\@eqnlabel{#1}}
\def\@eqnlabel{}
\def\@vacuum{}
\def\draftmarginnote#1{\marginpar{\raggedright\scriptsize\tt#1}}

\def\draft{\oddsidemargin -.5truein
        \def\@oddfoot{\sl preliminary draft \hfil
        \rm\thepage\hfil\sl\today\quad\militarytime}
        \let\@evenfoot\@oddfoot \overfullrule 3pt
        \let\label=\draftlabel
        \let\marginnote=\draftmarginnote
   \def\@eqnnum{(\theequation)\rlap{\kern\marginparsep\tt\@eqnlabel}%
\global\let\@eqnlabel\@vacuum}  }

\makeatletter
\newdimen\normalarrayskip              
\newdimen\minarrayskip                 
\normalarrayskip\baselineskip
\minarrayskip\jot
\newif\ifold             \oldtrue            \def\new{\oldfalse}
\def\arraymode{\ifold\relax\else\displaystyle\fi} 
\def\eqnumphantom{\phantom{(\theequation)}}     
\def\@arrayskip{\ifold\baselineskip\z@\lineskip\z@
     \else
     \baselineskip\minarrayskip\lineskip2\minarrayskip\fi}
\def\@arrayclassz{\ifcase \@lastchclass \@acolampacol \or
\@ampacol \or \or \or \@addamp \or
   \@acolampacol \or \@firstampfalse \@acol \fi
\edef\@preamble{\@preamble
  \ifcase \@chnum
     \hfil$\relax\arraymode\@sharp$\hfil
     \or $\relax\arraymode\@sharp$\hfil
     \or \hfil$\relax\arraymode\@sharp$\fi}}
\def\@array[#1]#2{\setbox\@arstrutbox=\hbox{\vrule
     height\arraystretch \ht\strutbox
     depth\arraystretch \dp\strutbox
     width\z@}\@mkpream{#2}\edef\@preamble{\halign
\noexpand\@halignto
\bgroup \tabskip\z@ \@arstrut \@preamble \tabskip\z@ \cr}%
\let\@startpbox\@@startpbox \let\@endpbox\@@endpbox
  \if #1t\vtop \else \if#1b\vbox \else \vcenter \fi\fi
  \bgroup \let\par\relax
  \let\@sharp##\let\protect\relax
  \@arrayskip\@preamble}
\def\eqnarray{\stepcounter{equation}%
              \let\@currentlabel=\theequation
              \global\@eqnswtrue
              \global\@eqcnt\z@
              \tabskip\@centering
              \let\\=\@eqncr
 \halign to \displaywidth\bgroup
    \eqnumphantom\@eqnsel\hskip\@centering
    $\displaystyle \tabskip\z@ {##}$%
    \global\@eqcnt\@ne \hskip 2\arraycolsep
         $\displaystyle\arraymode{##}$\hfil
    \global\@eqcnt\tw@ \hskip 2\arraycolsep
         $\displaystyle\tabskip\z@{##}$\hfil
         \tabskip\@centering
    &{##}\tabskip\z@\cr}
\begingroup\ifx\undefined\newsymbol \else\def\input#1 {\endgroup}\fi
\newfont{\hr}{msbm10}
\newfont{\ams}{msam10}

\font\numbers=cmss12
\font\upright=cmu10 scaled\magstep1
\def\stroke{\vrule height8pt width0.4pt depth-0.1pt}
\def\topfleck{\vrule height8pt width0.5pt depth-5.9pt}
\def\botfleck{\vrule height2pt width0.5pt depth0.1pt}
\def\Zmath{\vcenter{\hbox{\numbers\rlap{\rlap{Z}\kern 0.8pt\topfleck}\kern
2.2pt
                   \rlap Z\kern 6pt\botfleck\kern 1pt}}}
\def\Qmath{\vcenter{\hbox{\upright\rlap{\rlap{Q}\kern
                   3.8pt\stroke}\phantom{Q}}}}
\def\Nmath{\vcenter{\hbox{\upright\rlap{I}\kern 1.7pt N}}}
\def\Cmath{\vcenter{\hbox{\upright\rlap{\rlap{C}\kern
                   3.8pt\stroke}\phantom{C}}}}
\def\Rmath{\vcenter{\hbox{\upright\rlap{I}\kern 1.7pt R}}}
\def\Z{\ifmmode\Zmath\else$\Zmath$\fi}
\def\Q{\ifmmode\Qmath\else$\Qmath$\fi}
\def\N{\ifmmode\Nmath\else$\Nmath$\fi}
\def\C{\ifmmode\Cmath\else$\Cmath$\fi}
\def\R{\ifmmode\Rmath\else$\Rmath$\fi}

\newcounter{app}

\def\app{\setcounter{equation}{0}
\def\theequation{\Alph{app}.\arabic{equation}}\par
   \addvspace{4ex}
   \@afterindentfalse
  \secdef\@app\@dapp}

\newcommand\@app{\@startsection {app}{1}{0ex}%
                                   {-3.5ex \@plus -1ex \@minus -.2ex}%
                                   {2.3ex \@plus.2ex}%
                                   {\normalfont\Large\bf}}
\def\@dapp#1{%
{\parindent \z@ \raggedright  \bf #1}\par\nobreak}
\def\l@app#1#2{\ifnum \c@tocdepth >\z@
    \addpenalty\@secpenalty
    \addvspace{1.0em \@plus\p@}%
    \setlength\@tempdima{8em}%
    \begingroup
      \parindent \z@ \rightskip \@pnumwidth
      \parfillskip -\@pnumwidth
      \leavevmode \bfseries
      \advance\leftskip\@tempdima
      \hskip -\leftskip
      #1\nobreak\hfil \nobreak\hb@xt@\@pnumwidth{\hss #2}\par
    \endgroup\fi}
\newcounter{sapp}[app]

\def\sapp{\def\theequation{\Alph{app}.\arabic{equation}}
\par
\@afterindentfalse
  \secdef\@sapp\@dsapp}
\newcommand{\@sapp}{\@startsection{sapp}{2}{\z@}%
                                     {-3.25ex\@plus -1ex \@minus -.2ex}%
                                     {1.5ex \@plus .2ex}%
                                     {\normalfont\large\bfseries}}

\def\@dsapp#1{%
{\parindent \z@ \raggedright  \bf #1
}\par\nobreak}
\newcommand{\l@sapp}{\@dottedtocline{2}{1.5em}{2.3em}}

\def\d{\partial}

\def\bea{\begin{eqnarray}}
\def\eea{\end{eqnarray}}

\def\beq{\begin{equation}}
\def\eeq{\end{equation}}
\def\ba{\beq\new\begin{array}{c}}
\def\ea{\end{array}\eeq}
\def\be{\ba}
\def\ee{\ea}
\def\stackreb#1#2{\mathrel{\mathop{#2}\limits_{#1}}}
\def\Tr{{\rm Tr}}

\def\2{{1\over 2}}
\def\N2{${\cal N}=2$}
\def\4N{${\cal N}=4$}
\def\1N{${\cal N}=1$}

\begin{document}


\begin{flushright}
FIAN/TD-19/00\\
ITEP-TH-66/00\\
hep-th/0011222
\end{flushright}
\vspace{0.5cm}
\begin{center}
{\LARGE \bf Seiberg-Witten Toda Chains and N=1 SQCD
\footnote{Based on the talks
given at the {\sl E.S.Fradkin Memorial Conference, Moscow, June
2000 } and NATO workshop {\sl Integrable Hierarchies and Modern
Physics Theories, Chicago, July 2000 }.}}
\\
\vspace{1.0cm}
{\Large A.Marshakov
\footnote{e-mail
address: mars@lpi.ru,\ mars@gate.itep.ru,\ marshakov@nbivms.nbi.dk}}\\
\vspace{0.5cm}
{\it Theory
Department, Lebedev Physics Institute, Moscow ~117924, Russia \\
and \\ ITEP, Moscow 117259, Russia}, \\
\end{center}
\bigskip\bigskip
\begin{quotation}
We consider the Seiberg-Witten Toda chains arising in the context
of exact solutions to \N2 SUSY Yang-Mills and their relation to
the properties of \1N SUSY gauge theories. In particular, we
discuss their "perturbative" and "solitonic" degenerations and
demonstrate some relations of the latter ones to the confining
properties of
\1N vacua.
\end{quotation}

\section{Introduction}

The \N2 SUSY gauge theories possess continuous parametric families
of vacua, labeled by VEV's of complex scalar fields $\Phi $. An
exciting result is that the masses of BPS states and low-energy
effective actions for massless fields in these theories can be
found exactly \cite{SW} and are described in terms of integrable
systems so that the moduli of \N2 theories are identified with
moduli of complex structure of the corresponding spectral curves
\cite{SW,SW+} or integrals of motion of completely integrable
models \cite{GKMMM}. The corresponding family of integrable models
in the most simple case consists of the Toda chains \cite{Toda} of
exponentially interacting particles.

These \N2 SUSY gauge theories are even relatively realistic from
the point of view of their physical spectra, containing
propagating massless abelian gauge fields, but, as a consequence
of extended supersymmetry they possess also massless scalars --
the excitations of moduli, whose presence contradicts to what one
should expect in reality. The situation is seriously improved when
\N2 SUSY is violated down to \1N, then in contrast to \N2
situation, in \1N case a superpotential $W(\Phi )$ may be
generated which fixes the VEV of scalars to the extrema $dW=0$ and
one ends up with some restricted set (usually {\em finite} number)
of vacua where there is no more place for light moduli scalar
excitations. It is widely believed that \1N gauge theories are
already almost the theories close to real QCD, therefore
investigation of the properties of these vacua is an important
problem.

Below, we will try to present some arguments that the
Seiberg-Witten integrable systems, arose originally and associated
usually to the \N2 situation, may also play some role in
investigating properties of \1N gauge theories. We will start with
generic properties of SUSY algebra in supersymmetric Yang-Mills
theories and demonstrate that already at this level one may expect
appearance of certain hamiltonian and/or integrable structures.
Then we will pass to direct investigation of particular limits of
the Seiberg-Witten integrable systems, since it is clear from the
above reasoning that more realistic \1N picture appears only after
some conditions on moduli are imposed. We will discuss, in
particular, {\em  degenerations} of the smooth Seiberg-Witten
curves and demonstrate that, in fact, these degenerations can be
separated into two different classes, the properties of integrable
systems in these limits were studied in
\cite{BraMa}. It turns out, that the
structure of an integrable system implies a presence of a kind of
"regularizing" multiplet of {\em fundamental} matter -- to be
identified with the set of Baker-Akhiezer functions of an
integrable model, which can be explicitly constructed and their
properties will be discussed below. In particular, we will see
that the perturbative Baker-Akhiezer functions satisfy the
condition, which may be thought of as a (holomorphic square root
of) the D-term condition for the fundamental matter. We complete
our discussion by few remarks about the "internal" properties of
the Toda chain solutions.

\section{Supersymmetry and integrable structures}

The basic understanding of integrable structure in \N2 SUSY gauge
theories appears already at the level of the (extended) SUSY
algebra. Indeed \N2 SUSY algebra in 4D is generated by the
supercurrents $Q_A = \int d^3x J_A^0$ ($A=1,2$) and their complex
conjugated. These currents differ from each other by exchanging
the fermions $\{\psi_A\}$ and have general structure
\be
\label{susycu}
J_A^0 = (\psi_A F) \oplus (\psi_A {\bf D}) \oplus
\epsilon_{AB}(\bar\psi_B D\Phi)
\oplus \epsilon_{AB}(\psi_B{\bf F})
\ee
where $F=DA$ is gauge field strength, ${\bf
D}^a(\Phi^\dagger,\Phi) = [\Phi^\dagger,\Phi]^a$ is the D-term and
${\bf F}^a(\Phi) = {\d W\over\d\Phi^a}$ is the F-term
\footnote{ The superpotential terms are, in fact,
absent in \N2 theory, but they arise, in general, after breaking
SUSY down to \1N, which is important in the discussion below.}.
From this form and canonical commutation relations it is clear
that the anticommutator of $Q_1$ and $Q_2$ would contain the
following contributions
\be
\{ Q_1,Q_2\} = \sum_{A=1,2}
\{ (\bar\psi_A D\Phi) , (\psi_A F)\} +
\dots
\ee
and these terms lead to the central charges \cite{WO}
\be
Z_{SUSY} = {1\over g_{YM}^2}\int D\Phi\wedge (F + *F)
\ee
In the case of point-like charged objects the 3-dimensional
integral in the last formula can be rewritten using the Stokes
theorem as an integral over sphere ${\bf S}^2$ at spatial
infinity, giving rise to $Z_{SUSY} \sim a(n_e + \tau n_m)$, where
$a \sim \langle\Phi\rangle$ is VEV of scalar and $\tau = \vartheta
+ {i\over g_{YM}^2}$ is (complexified) coupling constant. However,
if we originally consider instead of ${\bf R}^3$ a compactified
space ${\bf R}^2\times{\bf S}^1$, the 2-dimensional integral is
rather taken over ${\bf S}^1\times {\bf S}^1$ and can be presented
as
\be
Z_{SUSY} = \int_{{\bf S}^1\times{\bf S}^1}  d^2x
\epsilon^{ij}\d_i\Phi \d_j A = \int_{2-cycle}
\delta\Phi\wedge\delta A
\ee
The last integral may be, in fact, identified with the integral of
{\em holomorphic symplectic} form $\delta\Phi\wedge\delta A$ where
the appearance of extra moduli parameters $A$ is related to the
Wilson-Polyakov loops of gauge fields, complexified by dual
photons and becomes clear if one considers compactification of 4D
theory down to 3D plus possible one compact dimension \cite{SW3}.
The loops of gauge field break the invariance in colour space and,
in particular, this leads to replacement of the diagonal matrix of
the VEV $\Phi$ by, in the simplest case of Toda chains "almost
diagonal", Lax operator ${\cal L}$ of an integrable model.

Moreover, in the compactified case one may violate \N2 4D SUSY by
the Scherk-Schwarz mechanism, which leads to generation of
superpotentials, in spirit of \cite{AHW}. These superpotentials
exactly coincide with the Hamiltonians of the Seiberg-Witten
integrable systems \cite{AHW,WeqH}. It means, in particular, that
integrable models are even more visible in {\em compactified}
\1N theory (\N2 in 3D terms), than in 4D \N2 case.

\begin{figure}[tp]
\epsfysize=4cm
\centerline{\epsfbox{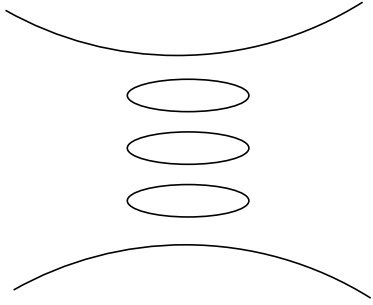}}
\caption{\sl  Smooth Seiberg-Witten curve for the \N2 $SU(N)$ pure gauge theory
associated with periodic Toda chain.}
\label{fi:smooth}
\end{figure}

\begin{figure}[tp]
\epsfysize=8cm
\centerline{\epsfbox{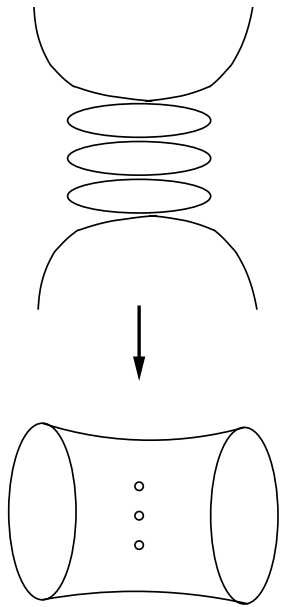}}
\caption{\sl  A perturbative degeneration of the Seiberg-Witten curve
from fig.~\ref{fi:smooth} shrinks into its solitonic limit.}
\label{fi:persol}
\end{figure}

SUSY breaking from \N2 down to \1N (in 4D terms) is usually
associated with {\em degeneration} \cite{SW,DouShe,GVY} of smooth
Seiberg-Witten curve for \N2, see fig.~\ref{fi:smooth}.
In fact not {\em any} degeneration of a curve corresponds to
\1N vacuum. Roughly speaking there are two classes of degenerations,
the "open"  or {\em perturbative} and {\em solitonic}, if speaking
in the language of integrable systems, the same may be thought of
as asymptotically free and zero-charge from the point of view of
corresponding quantum field theories. Only the latter class of
(solitonic or zero-charge) degenerations may play the role of set
of vacua for \1N SUSY gauge theories. Qualitatively the difference
is clearly seen in the brane picture for the Seiberg-Witten Toda
chain curves \cite{WittM,MMaM,WittMQCD,HaZaS}, the brane with {\em
positive} tension would like to "shrink" to the zero-charge
points, even if you start from different, perturbative
degeneration (see fig.~\ref{fi:persol}).

\section{Integrable Toda chains from weak coupling}

We start from {\em perturbative} degeneration, where
\N2 SUSY effective actions and integrable systems behind are completely
determined by the 1-loop contributions. In the example of pure \N2
gauge theory with $SU(N)$ gauge group \cite{SW+}, the scalar field
$\Phi $ acquires nonzero VEV's $\Phi = {\rm diag}(a_1,\dots,a_N)$
extremizing the potential $\Tr [\Phi,\Phi^\dagger ]^2 $. This
(generically) breaks the $SU(N)$ gauge group due to the
interaction term $\left([A_\mu,\Phi]_{ij}\right)^2 =
\left(A_\mu^{ij}( a_i- a_j)\right)^2$ and the masses of $W$-bosons
(and their superpartners) are classically given by $ a_{ij}\equiv
a_i- a_j$ or
\be
\label{pertsample}
a_{ij} =
\oint_{C_{ij}}dS^{\rm pert} = \oint_{C_{ij}} \lambda d\log w=
\oint_{C_{ij}} \lambda d\log P_{N}(\lambda )
\ee
for a particular ``figure-of-eight" like contour $C_{ij}$ around
the roots $\lambda =  a_i$ and $\lambda =  a_j$ of the generating
polynomial
\be
w = P_N(\lambda ) = \det_{N\times N} (\lambda - \Phi) =
\prod_{i=1}^N (\lambda -  a_i),
\label{polyn}
\ee
This means that the contour integrals in the complex $\lambda
$-plane (\ref{pertsample}) for such contours around these singular
points (the roots of the polynomial (\ref{polyn})) give one set of
the BPS masses in the Seiberg-Witten theory. Another set of masses
(monopoles) may be associated to {\em dual} contours starting and
ending at the points $a_i$ where $dS^{\rm pert}$ has
singularities, taking into account that divergences are absorbed
by a renormalization of the couplings. The integrals over the
contours on the marked plane are in fact contour integrals on a
{\em degenerate} Riemann surface depicted at the top of
fig.~\ref{fi:persol}: if one cuts it (vertically) into two pieces,
connected by very narrow tubes, one gets exactly two copies of the
desired $\lambda$-plane.

The \N2 perturbative effective action (or prepotential) ${\cal F}$
and the set of effective couplings $-i\pi T_{jk} = {\d^2{\cal
F}\over\d a_j\d a_k}$, are given by the 1-loop diagrams with the
result
\footnote{Instanton contributions to the prepotential which are proportional
to powers of $\Lambda^{2N}$ are suppressed, and one keeps only the
terms proportional to $\log\Lambda$ or the bare coupling $\tau $.}
\be
\label{effcharge}
 T_{jk}^{\rm pert} = {1\over
2\pi i}\sum_{({\rm masses M})_{jk}}
 \log {({\rm mass})^2\over\Lambda ^2} =
{1\over 2\pi i}\log {a_{jk}^2\over\Lambda ^2}
\ee
where the scale parameter $\Lambda\equiv\Lambda_{QCD}$ so
introduced may be related to the bare coupling $\tau $.

The connection with integrable systems comes by interpreting the
differential $dS^{\rm pert}$ with the generating differential of a
Hamiltonian system -- an open Toda chain or Toda molecule
\footnote{From the simplest $SU(2)$ pure gauge theory
\cite{SW},  with the curve (\ref{polyn}) $w = \lambda^2 - h$,
$h = \2\Tr\Phi^2= a^2$ it becomes clear, that the masses
(\ref{pertsample}) are defined by the contour integrals of
$dS^{\rm pert} = {\lambda d\lambda\over\lambda - \sqrt{h}} +
{\lambda d\lambda\over\lambda + \sqrt{h}}$, which can be
interpreted as an {\em integration} of open Toda-Liouville model
with Hamiltonian $h =p\sp2+e\sp{2q}$, provided $\lambda\to p$,
$w\to e^{2q}$. The same is true in general, that is, integration
of the canonical differential $dS =2 pdq$ over the trajectories of
the Toda chain solutions gives rise, for various (complexified)
trajectories, to the BPS masses in the Seiberg-Witten theory
\cite{GKMMM}.}. Remarkably, this interpretation extends beyond
the perturbative regime, the "regularized" curve for (\ref{polyn})
(see fig.~\ref{fi:smooth}!)
\be
\label{fsc-Toda}
w + {\Lambda^{2N}\over w} = P_N(\lambda )
\ee
is the spectral curve of a {\em periodic} Toda chain (an analog of
the Sine-Gordon model in the same sense as the Toda molecule is a
multiparticle analog of the Liouville model) providing the {\em
exact} non-perturbative solution taking into account all
instantonic contributions. The residue formulas (\ref{pertsample})
for the smooth curve (\ref{fsc-Toda}) of genus $N-1$ turn into the
action period integrals
\be
({\bf a}, {\bf a}^D)  = \oint_{{\bf A},{\bf B}}\lambda{dw\over w}
\ee
so that the set of effective charges (\ref{effcharge}) is
generalized to the {\em period matrix} of (\ref{fsc-Toda})
\be
T_{ij} = {\d a^D_i\over\d a_j} = {\d^2{\cal F}\over\d a_i\d a_j}
\ \ \ \ \ \ \ i,j=1,\dots,N-1
\ee
where ${\cal F}$ now stands for {\em non-perturbative}
prepotential or effective action.

\section{More on Toda: periodic chain and molecule}

The dynamics of the integrable system itself may be again seen
only in the compactified version \cite{SW3} when the number of
moduli of the gauge theory jumps up to the whole phase space of an
integrable model. In this case the angle variables play the role
of extra moduli related directly to the hamiltonian "times". This
relation is especially simple in perturbative regime, when the
basis of (normalized) "holomorphic" differentials
\be
2\pi id\omega_i = {d\lambda\over\lambda-a_i}
\ee
and corresponding angle variables (originally defined as
co-ordinates in the Jacobian of smooth curve (\ref{fsc-Toda}) of
fig.~\ref{fi:smooth}) are
\be
\label{zlambda}
2\pi iz_i = \sum_{k=1}^N\int_{P_0}^{P_k}{d\lambda\over\lambda-a_i}
=
\sum_{n=0}^{\infty} a_i^n\sum_{k=1}^N\int_{P_0}^{P_k}
{d\lambda\over\lambda^{n+1}}\
\\
\stackreb{\lambda_k\to\lambda(P_0)=\infty}{\sim}\
N\log\lambda + \sum_{n=1}^{\infty}a_i^nt_n
\ee
for some divisor $\{ P_k \}$ with $\lambda_k\equiv\lambda(P_k)$
and
\be
\label{miwa}
t_n = - {1\over n}\sum_{k=1}^N\lambda_k^{-n}
\ee
are the "hamiltonian" times in Miwa variables. It means, that
after renormalization $z_i\to z_i - {N\over 2\pi i}\log\Lambda$
one gets
\be
2\pi iz_i = a_it + \sum_{n>1}a_i^nt_n
\ee
A natural ingredient of the integrable formulation is  the set of
the Baker-Akhiezer functions $\{ \Psi_i \}$ (and
$\{\tilde\Psi\}$), which can be thought of as introducing of
"regularizing" fundamental matter. Like quark multiplets in SUSY
gauge theories the Baker-Akhiezer functions always appear "in
pairs" and carry a single index of lowest fundamental, ${\bf N}$
-- for $\{ \Psi_i \}$, or  $\bar{\bf N}$ -- for $\{\tilde \Psi_i
\}$, representation of gauge group. Again, in 4D theory they are,
up to gauge transformation, just the eigenvectors of diagonal
matrix $\Phi$, but in compactified version, when $\Phi$ should be
replaced by the Lax operator ${\cal L}$, they become nontrivial.

On degenerate spectral curve (\ref{polyn}) the set of the
Baker-Akhiezer functions can be defined by the following analytic
requirements \cite{BraMa}:
\begin{itemize}
\item This is the set of functions $\Psi_k=\Psi_k(\lambda )$,
$k=0,\dots,N-1$
\footnote{The $N$-dimensional vectors like $\{\Psi_k\}$
are labeled here for simplicity of notations as $k=0,\dots,N-1$,
unlike usual conventions in group theory.} which have exactly $k$
zeroes on rational curve (\ref{polyn}) and a single pole of order
$k$ at $\lambda=\infty$. For $k\geq N$ they can be defined by
$\Psi_{k+N}=w\Psi_k$.
\item These functions can be constructed as linear combinations of
$\prod_{i_1<\dots <i_k}(\lambda-a_{i_l}) = \prod_{j\in
I}(\lambda-a_j) $ or, vice versa, certain linear combinations of
$\Psi_k$ vanish exactly at the chosen sets $I$ of $k$ points
$a_{i_1},\dots,a_{i_k}$, $i_1<\dots <i_k$. One can see that the
Baker-Akhiezer functions, indeed, satisfy for $e^{Z_i}
\equiv e^{2\pi i z_i}{\prod_{j\neq i}}{1\over a_{ij}}$
\be
\label{gluesum}
\sum_{i=1}^Ne^{Z_i}\Psi_k(a_i)  =
(-)^{k+1}\sum_{|I|=k+1}\prod_{i\in I}e^{Z_i}\prod_{i,j\in I}a_{ij}
= 0
\ee
\end{itemize}

It is easy to see, that so defined Baker-Akhiezer functions are
related to the tau-functions and can be written (up to
normalization) as
\be
\label{ba}
\Psi_k(\lambda) = \lambda^k\ \Theta_k\left( z_i - \sum_n{a_i^n\over
n\lambda^n} \right) = {\sum}^k \prod_{i\in I}e^{2\pi iz_i}(\lambda
- a_i){\prod_{j\in{\bar I}}} {1\over a_{ij}}
\ee
where we write all $a_{ij}$ in such order that $i<j$. The sum
${\sum}^k$ is taken over all sets of $|I|=k$ indices with $\bar I$
denoting the complementary set to $I$. The tau-functions
\be
\label{thetaop}
\Theta_k
= \sum_{\sum_{i=1}^Nn_i=k;\ \sum_{i<j} (n_i-n_j)^2=k(N-k)} e^{2\pi
i\sum_{j=1}^Nn_j z_j - i\pi\sum_{i<j=1}^N(n_i-n_j)^2T_{n_in_j} } =
\\
 = {\sum}^k \prod_{i\in I}e^{2\pi iz_i}{\prod_{j\in{\bar I}}}
{1\over a_{ij}}
\ee
are nothing but degenerate theta-functions for the Toda molecule
system where in the perturbative limit $\Lambda\to 0$ the period
matrix $T_{ij}$ is given by (\ref{effcharge}) with the terms
proportional to $\log\Lambda $ absorbed  into some
renormalizations:
 $z_i\to  z_i-{N\over 2\pi i}
\log\Lambda$, $\Theta_k\to\Lambda^{k^2}\Theta_k$
\footnote{The last renormalization could be understood as a logarithmic
renormalization of the bare coupling $\tau $.}. They can be
obtained as degenerations of the theta-functions on (Jacobian of)
smooth curve (\ref{fsc-Toda}) and their properties can be found,
for example, in \cite{BraMa}. The formulas (\ref{ba}) can be
written down even more explicitly, say for the $N=2$ case one gets
\be
\label{psi1n2}
\Psi_1(\lambda ) = {\lambda\over a}\left(\exp\left(z-\sum_n{a^n\over n\lambda^n}\right)
+\exp\left(-z - \sum_n(-)^n{a^n\over n\lambda^n}\right)\right) =
\\
= {e^z+e^{-z}\over a}\left(\lambda - a\tanh z\right) =
e^{-q}(\lambda - p)
\ee
providing the relations $\Theta_1 = {1\over a}(e^z + e^{-z}) =
e^{-q}$ and $p=a\tanh z$
\footnote{Note, that $\Psi_2=\lambda^2-a^2=0$ when $\lambda=\pm a$ belongs for a spectrum
of the Toda molecule. This is a common feature for all
$\Psi_N=\prod_{i=1}^N(\lambda-a_i)$.}. One may also easily check
that the vector
\be
\Psi(\lambda ) = \left(
\begin{array}{c}
  1 \\
  \Psi_1(\lambda)
\end{array}\right)
\ee
is an eigenvector of the open Toda Lax operator ${\cal L}\Psi =
\lambda\Psi$
\be
{\cal L} =
\left(
\begin{array}{cc}
  p & e^q \\
  e^q & -p
\end{array}\right)
\ee
provided by $\lambda^2 = p^2 + e^{2q} = h = a^2$, which is related
to the VEV of 4D scalar $\Phi$ by ${\cal L}=U\Phi U^{-1}$, where
matrix $U$ is constructed from the Baker-Akhiezer functions.

The {\em dual} Baker-Akhiezer functions
$\{\tilde\Psi_k(\lambda)\}$ (which may be thought as coming from
the "dual" sheet of degenerate Toda chain curve \cite{KriVa}) can
be written, up to normalization, again in terms of the
theta-functions as
\be
\label{badagg}
\tilde\Psi_{N-k-1}(\lambda) = \lambda^{-k}\ \Theta_k\left( z_i + \sum_n{a_i^n\over
n\lambda^n} \right) = {\sum}^k \prod_{i\in I}{e^{2\pi
iz_i}\over\lambda - a_i}{\prod_{j\in{\bar I}}} {1\over a_{ij}}
\ee
For example
\be
\tilde\Psi_{N-1}= {1\over\prod_{i=1}^N(\lambda - a_i)}
\\
\tilde\Psi_{N-2}(\lambda) = \lambda^{-1}\ \Theta_1\left( z_i + \sum_n{a_i^n\over
n\lambda^n} \right) =
\sum_{i=1}^N {e^{2\pi iz_i}\over\lambda - a_i}{\prod}'{1\over a_{ij}} =
{\sum_{i=1}^N e^{2\pi iz_i}{\prod}'_{j\neq i}{\lambda-a_i\over
a_{ij}}\over
\prod_{i=1}^N(\lambda-a_i)}
\\
\dots
\ee
so that for the $N=2$ case one gets
\be
\label{psi1dn2}
\tilde\Psi_1(\lambda )={1\over\lambda^2-a^2}
\\
\tilde\Psi_0(\lambda ) = \lambda^{-1}\ {1\over a}\left(
e^{z+\sum_n{a^n\over n\lambda^n}} +e^{-z+ \sum_n(-)^n{a^n\over
n\lambda^n}}\right) =
 {{e^z\over\lambda-a}+{e^{-z}\over\lambda+a}\over a} =
\\
= \left({e^z+e^{-z}\over a}\right)\ {\lambda + a\tanh
z\over\lambda^2-a^2} = e^{-q}{\lambda + p\over\lambda^2-a^2}
\ee
It is easy to see, for (\ref{psi1n2}) and (\ref{psi1dn2}) in
particular, that one may indeed impose {\em gluing} conditions
\be
\label{glupert}
\Psi_i(a) = C\tilde\Psi_i(a)
\ee
for some coefficient $C$. In our basis (\ref{badagg}) for the dual
Baker-Akhiezer functions this coefficient has to be nontrivial,
since $\tilde\Psi_k(a)$ are naively singular exactly in the points
of the spectrum, but adjusting $C$, or changing the normalization
of (\ref{badagg}) we see, that the gluing condition
(\ref{glupert}) indeed holds.

These are the gluing conditions at weak coupling -- and they may
be thought of as taking holomorphic square root from the D-term
condition for a single flavour of "regularizing" fundamental
matter.

\section{The strong coupling limit}

The quantum moduli space of the 4D pure $SU(N)$ \N2 SYM has $N$
maximally singular points at which $N-1$ monopoles become
simultaneously massless -- the confining vacua of an \1N theory
\cite{SW,SW+,DouShe,HaZaS}, where the dual variables
$a_i\sp{D}= {\d{\cal F}\over\d a_i}$ are appropriate to describe
the prepotential. The \N2 spectral curve (\ref{fsc-Toda}) at these
points is described in terms of a Chebyshev polynomial (in this
section we put $\Lambda=1$)
\be\label{chebyshev}
P_N^{\rm Chebyshev}(\lambda)= 2\cosh z.
\ee
with a solution $\lambda=2\cosh\left({z\over N}+i{2\pi k\over
N}\right)$ ($k=0,\ldots N-1$). These are the \1N points of the
theory, related by a ${\bf Z}_N$ symmetry. The hyperelliptic form
of the spectral curve (\ref{fsc-Toda}) (recall $y=w-1/w$) is now
\be
\label{solpeto}
y^2 = P_N(\lambda)^2 - 4 = (\lambda^2-4)
\prod _{j=1}^{N-1}(\lambda - 2\cos{\pi j\over N})^2
\ee
which is a ``solitonic" curve in the {\em periodic} Toda chain
\cite{AMMinn,BraMa}: now the corresponding $\bf B$ periods have collapsed,
in contrast to the perturbative case. Introducing
\begin{equation}
\label{mapcyl}
\lambda = 2\cosh {z\over N} \equiv \xi + \xi ^{-1}.
\end{equation}
one maps the 2-sheeted cover of the $\lambda$-plane
$y=\sqrt{\lambda^2-4}$ to a cylinder with co-ordinate $\xi$. Thus
eqs.~(\ref{chebyshev}), (\ref{solpeto}) describe analytically a
cylinder with $N-1$ distinguished pairs of points -- an inside-out
S-duality transform of the bottom curve from fig.~\ref{fi:persol}.
It is worth noting that in contrast to perturbative regime (the
upper curve on fig.~\ref{fi:persol}) now a curve does {\em not}
separate into two pieces. The canonical basis of "holomorphic"
differentials in this limit can be can be chosen as
($j=1,\ldots,N-1$)
\be
d\omega^D_j = {\sin{\pi j\over N}\over\pi}{d\xi\over
\left(\xi - e^{i\pi j\over N}\right)\left(\xi - e^{-{i\pi j\over N}}\right)}
\ \ \stackreb{\xi_j^\pm\equiv e^{\pm{i\pi j\over N}}}{=}\ \
{1\over 2\pi i}d\log{\xi - \xi_j^+\over\xi - \xi_j^-}.
\ee
These differentials are normalized to the ${\bf B}$-cycles, here
the cycles around the marked points $\xi_j^\pm$,
\be
\oint_{B_i}d\omega^D_j =  \oint_{\xi_i^+}d\omega^D_j
=- \oint_{\xi_i^-}d\omega^D_j = \delta_{ij},
\ee
while certain of the ${\bf A}$-periods ($\oint_{A_j}d\omega^D_j =
\int_{\xi_j^-}^{\xi_j^+}d\omega^D_j$) diverge, the others ($j\ne k$)
being given by
\footnote{This formula was independently proposed in \cite{EMMa}.}
\be
\label{invt}
T^D_{jk} = \oint_{A_j}d\omega^D_k = {1\over 2\pi
i}\log{\sin^2{\pi\over 2N}(j-k)\over
\sin^2{\pi\over 2N}(j+k)} =
{1\over i\pi}\log{\sin{\pi\over 2N}|j-k|\over
\sin{\pi\over 2N}(j+k)}.
\ee
This is an exact formula for the set of effective couplings in
strong-coupling regime, which has "typical solitonic" form. The
appearance of solitons at strong coupling is a sort of "intuitive
explanation" of the absence of the point with
$\langle\Phi\rangle=0$ since the solitonic points correspond, on
one hand, to maximal degeneracy and, on the other hand, to minimum
of superpotential, being an analog of the "BPS" first-order
solutions in Seiberg-Witten Toda chains.

The angles at strong coupling are given by
\be
\label{abelsol}
z_j^D = \sum_{k=1}^{N-1}\int^{\xi_k} d\omega_j^D = {1\over 2\pi i}
\sum_{k=1}^{N-1}\log{\xi_k - \xi_j^+\over\xi_k - \xi_j^-} =
{1\over\pi}\sum_{n=1}^\infty t_n\sin{\pi jn\over N}
\ee
where $t_n = \sum_k{\xi_k^{\pm n}\over n}$ depending on whether
$\{\xi_k\}$, $k=1,\dots,N-1$ go to $0$ or to $\infty$. We observe
also that the relation (\ref{abelsol}) coincides with the vacuum
value of the string tension, or monopole condensate, proportional
to the SUSY breaking parameters $t_k$ \cite{DouShe,HaZaS,AMMinn,EMM}.
The detailed study of the properties of these strings are beyond the
scope of this note, a discussion of related questions can be found in
\cite{EFMG}.

In contrast to the weak-coupling regime, the gluing conditions at
the strong coupling are \cite{AMMinn,BraMa}
\be
\Psi_n (\xi_j^+) = \Psi_n(\xi_j^-)\ \ \ \ \ \ \ j=1,\dots,N-1.
\label{glue}
\ee
which means that the Baker-Akhiezer function on a cylinder
remembers that it originally came from a genus $N-1$ Riemann
surface, and each pair of points $\xi_j^+,
\xi_j^-={1\over\xi_j^+}$ corresponds to a degenerate handle. Note,
that the form of condition (\ref{glue}) is different from
(\ref{glupert}) (and in fact, is even more similar to
(\ref{gluesum})) and relates the same Baker-Akhiezer functions in
different points of the strong-coupling spectrum. There is no
natural relation between $\Psi$ and $\tilde\Psi$ in solitonic
phase, which should be probably related to the properties of
fundamental matter multiplets at strong coupling. We are going to
return to this problem elsewhere.

\section{More remarks on the Toda tau functions}

In this section we will try to make few more remarks concerning
the degenerate limits of the Toda chain, which do not have at the
moment direct relation to the properties of \1N SQCD, nevertheless
being of some interest by themselves. We believe, however, that
their physics origin is related to the bulk/boundary duality
between SUSY gauge theories and gravity.

{\bf Commutativity of the theta-functions}.

The theta-functions in (\ref{thetaop}) may be obtained as
degenerate limit of the coefficients $\Theta_{k}$ in the expansion
\cite{BMMM3} of the genus $N$ theta function
\be\label{dectoda}
\Theta_N  = \sum_{k\in{\Z }_N}
e^{2\pi i{k\over N}z} \Theta_{k}({\bf z}|T)
\ee
We may already deduce an interesting consequence: the ratios of
the coefficients $\Theta_k\equiv\Theta_{k}({\bf z}| T )$ Poisson
commute (see \cite{M99c} and references therein),
\be
\label{poicom}
\left\{ {\Theta_i\over\Theta_j}, {\Theta_{i'}\over\Theta_{j'}}\right\} = 0,
\qquad
\forall\ i,j,i',j'.
\ee
The Poisson bracket here is that corresponding to the symplectic
form
\be
\label{sympfor}
\Omega^{Toda} = \sum_{i=1}^{N-1}d z_i\wedge da_i =
\sum_{i=1}^{N-1}d q_i\wedge d p_i,
\ee
where $q_i$ and $ p_i$ are co-ordinates and momenta of the Toda
chain and we are working in the centre of mass frame. The Poisson
commutativity of the ratios (\ref{poicom}) follows from the
solution of the periodic Toda chain \cite{DaTa,KriTo}
\be
\label{todasol}
e^{q_i} = {\Theta_i\over\Theta_{i-1}},
\qquad
{\Theta_i\over\Theta_j} = \prod_{k=j+1}^i e^{ q_k}
\ee
(Here  $\Theta_0=\Theta_N$ and the $\Theta_k$ are the Toda chain
tau-functions depending on the discrete time $k$, the number of
the particle.) Now because the coordinates $ q_i$ obviously
Poisson commute, $\{  q_i, q_j\} = 0$, we deduce (\ref{poicom}).
This expression gives a precise formulation of the old expectation
that the Toda chain tau-functions Poisson commute with each other.
In the case of {\em periodic} Toda chain, however, only the {\em
ratios} are commutative (theta-functions themselves, co-ordinates
in projective space, are defined only up to common
multiplication), but in the degenerate perturbative limit
$\Theta_0=\Theta_N=1$ and, thus, tau-functions commute themselves.
Moreover, in the perturbative limit formulas (\ref{thetaop}) and
(\ref{todasol}) give an explicit solution to the Toda molecule
\cite{Toda} in terms of the action-angle variables.

{\bf Relation to matrix models}.

It is widely known now that the partition functions of the matrix
models are tau functions of Toda chains \cite{GMMMO}. Expression
(\ref{thetaop}), being a finite sum, still is a sort of partition
function, when the sum is taken over the discrete spectrum of the
finite Toda molecule. It is easy to demonstrate, that, indeed this
tau function can be considered as a "discretized" analog of the
partition function of hermitian 1-matrix model.

For example, one can verify that
\be
\label{moma1}
\Theta_k = \left.\det_{k\times k}K_{n+m}\right|_{n,m=1,\dots,k}
\ee
where the "moment matrix" $K_{nm}=K_{n+m}$ is "an average" w.r.t.
$\Theta_1$, i.e.
\be
\label{moma}
K_n = \langle a^n\rangle_1 = \sum_{i=1}^N e^{Z_i}a_i^{n-1}, \ \ \
\ \ \
\Theta_1 = \sum e^{Z_i}
\ee
(cf. with \cite{GMMMO,KMMOZ}). The simplest proof \cite{BraMa} is
based on the Cauchy-Binet formula for the $k\times N$  ($k\leq N$)
rectangular matrices $A_{ni}$ and $B_{im}$ ($i=1,\dots,N$,
$n,m=1,\dots,k$)
\be
\left.\det_{k\times k}\left(\sum_{i=1}^N A_{ni}B_{im}\right)
\right|_{n,m=1,\dots,k} =
\sum_{i_1<\dots <i_k}
\left.\det_{k\times k} A_{ni}\right|_{i\in I;\ n=1,\dots,k}
\left.\det_{k\times k} B_{im}\right|_{i\in I;\ n=1,\dots,k}.
\ee
so that
\be
\label{mamo}
\Theta_k =
{\sum}_{|I|=k} \prod_{i\in I}e^{2\pi i z_i}{\prod_{j\in{\bar I}}}
{1\over a_{ij}} = {\sum}_{|I|=k}\prod_{i\in I}e^{Z_i}{\prod_{i\ne
j\in I}}a_{ij} = {\sum}_{|I|=k}\prod_{i\in I}e^{Z_i}{\prod_{i<j;\
i,j\in I}}a_{ij}^2
\ee
is indeed a ``discretized" analogue of the tau-function of
1-matrix model \cite{KMMOZ} since for {\em any} coefficients $C_i$
\begin{equation}
\begin{array}{l}
{\sum}_{|I|=k}\prod_{i\in I}C_i\,{\prod_{i<j;\ i,j\in
I}}a_{ij}^2\equiv
\sum_{I:\ i_1<\dots <i_k}C_{i_1}\dots C_{i_k}\prod_{i_n<i_m}\left(a_{i_n}-
a_{i_m}\right)^2
\\
\qquad= \sum_{i_1<\dots <i_k}C_{i_1}\dots C_{i_k}
\left.\det_{k\times k} a_i^{n-1}\right|_{i\in I;\ n=1,\dots,k}
\left.\det_{k\times k} a_i^{m-1}\right|_{i\in I;\ m=1,\dots,k}
\\
\qquad= \left.\det_{k\times k}\left(
\sum_{i=1}^NC_ia_i^{n+m-2}\right)\right|_{n,m=1,\dots,k}.
\end{array}
\end{equation}
and, substituting $C_i=e^{Z_i}$, one arrives at
eqs.~(\ref{moma1}), (\ref{moma})
\be
\label{mamodet}
\Theta_k = \sum_{I:\ i_1<\dots <i_k}e^{Z_{i_1}+\dots Z_{i_k}}
\prod_{i_n<i_m}\left(a_{i_n}-
a_{i_m}\right)^2 =
\left.\det_{k\times k}\left(
\sum_{i=1}^Ne^{Z_i}a_i^{n+m-2}\right)\right|_{n,m=1,\dots,k} =
\det_{k\times k}K_{n+m}.
\ee
The "non-discretized" (though still "discrete") 1-matrix model
should be restored in the $N\to\infty$ limit of the Toda molecule,
under condition that the spectrum $\{ a_i \}$ fills in the whole
real axis. Then instead of the Cauchy-Binet formula (rectangular
matrices become infinite in one of the direction) one can
effectively use the technique of the orthogonal polynomials
\cite{GMMMO,KMMOZ}.

\section{Conclusion}

We have discussed just a brief list of questions related to the
problems of \1N vacua is SUSY gauge theories. However, already at
this level we can see that there are remarkable coincidence
between the properties of \1N theories and Seiberg-Witten Toda
chains "at strong coupling". They include the relations between
the Baker-Akhiezer functions and the formulas for the phases of
solitons, which look like being "in charge" for confining strings
in SUSY gauge theories. It is necessary also to point out that
perturbative integrable system is "stable" under small changes of
moduli $a_i$, while in strong coupling regime all moduli are
completely "confined" to certain values.

\section*{Acknowledgements}

I am indebted to H.Braden, I.Krichever, A.Vainshtein and A.Yung
for many illuminating discussions on different subjects considered
in this note and to V.Zaykin and A.Sorin for encouragement to
write this contribution. The work was partially supported by RFBR
grant No. 99-02-16122, INTAS grant No. 99-0590 and CRDF grant No.
RF-2102 (6531) and grant for the support of scientific schools No.
00-15-96566.

\end{document}